# A low cost remote sensing system using PC and stereo equipment

Joel F. Campbell, Michael A. Flood, Narasimha S. Prasad, and Wade D. Hodson
NASA Langley Research Center, MS 488, Hampton, Virginia 23681
joel.f.campbell@nasa.gov

## ABSTRACT

A system using a personal computer, speaker, and a microphone is used to detect objects, and make crude measurements using a carrier modulated by a pseudorandom noise (PN) code. This system can be constructed using a personal computer and audio equipment commonly found in the laboratory or at home, or more sophisticated equipment that can be purchased at reasonable cost. We demonstrate its value as an instructional tool for teaching concepts of remote sensing and digital signal processing.

## I. INTRODUCTION

The measurement of $CO_2$ concentrations in the atmosphere from a satellite using continuous wave (CW) light detection and ranging (LIDAR)[1] is of considerable interest. To test these methods in the lab, we implemented them using sonar. The advantages to using sonar include that the speed of sound is much slower than that of light, reasonable resolution can be achieved at relatively low bit rates, and a slow sampling system such as a sound card may be utilized. Our system has helped NASA summer students learn about remote sensing using equipment that is familiar to most of them. This equipment includes personal a computer, stereo amplifier, microphones, and speakers. Because many students are interested in music, this demonstration of remote sensing is one way for students to develop a greater interest in physics and engineering.

Traditional techniques for determining the distance to a target usually focus on transmitting a pulse and measuring the time to the target and back. The distance to the target is $d = c\ \tau / 2$, where c is the speed of sound and $\tau$ is the time to the target and back. The factor of two comes from the fact we must take into account the distance to the target and back, which is 2d.

Other techniques use a continuous wave modulated in some unique way. Signal processing is employed to determine the distance to the target. The idea is to transmit a repeating pattern for the modulation. To determine the distance, we need to compute the time for the pattern to reach the target and back.

Signal processing techniques may be employed to convert the pattern into a pulse. In our case we correlate the outgoing pattern with the incoming pattern. The position of that pulse is proportional to the target distance just as in standard pulse sonar.[1] There are many possible choices for the modulation pattern. A good pattern is one that produces a clean pulse without side lobes when the incoming and outgoing patterns are correlated. From a theoretical standpoint pseudorandom noise codes in the form of maximum length sequences are best because they produce the best pulse profile possible in the form of a delta function.

## II. MAXIMUM LENGTH SEQUENCES

Although other schemes are possible such as white noise, we used maximum length sequences as the main modulation scheme. Maximum length sequences have very good autocorrelation properties, which means that when the modulation pattern is correlated with itself, a single spike is produced without side lobes or other artifacts. For this reason they are very useful in LIDAR, radar, and sonar applications.[1,2] Because these sequences are digital, they are very easy to compute. Popular representations include linear feedback registers and generating polynomials. In this paper we use a simple recursion relation because it is the simplest way to generate these sequences in software.

To generate the individual bits/elements in the sequence we represent the recursion relation using exclusive-or, which we represent by the $\oplus$ symbol. If N1 and N2 are binary numbers represented by 0 or 1 then

$$\begin{aligned} N_1 \oplus N_2 &= 0, N_1 = N_2, \\ N_1 \oplus N_2 &= 1, N_1 \neq N_2. \end{aligned} \qquad (1)$$

An example of a sequence that repeats every 255 elements is

$$X_{j+8} = X_j \oplus X_{j+2} \oplus X_{j+5} \oplus X_{j+6}, \quad (2)$$

where $X$ is the maximum length sequence and $j$ is the index. For the initial values of $X_0$ - $X_7$ we may choose any sequence of 0s and 1s we wish as the seed as long as they are not all 0 or all 1. For the seed (1,0,1,0,1,1,1,1) we generate the repeating sequence

(1, 0, 1, 0, 1, 1, 1, 1, 0, 0, 1, 1, 1, 0, 1, 0, 0, 0, 0, 1, 0, 1, 0, 1, 1, 0, 0, 1, 0, 1, 0, 0, 0, 1, 0, 1, 1, 0, 0, 0, 0, 0, 1, 1, 0, 0, 1, 0, 0, 0, 1, 1, 0, 0, 0, 0, 1, 1, 0, 1, 1, 1, 1, 1, 0, 1, 1, 1, 0, 0, 0, 0, 1, 0, 0, 0, 0, 0, 1, 0, 0, 1, 0, 1, 0, 1, 0, 0, 1, 0, 1, 1, 1, 1, 1, 0, 0, 0, 0, 0, 0, 1, 1, 1, 0, 0, 1, 1, 0, 0, 0, 1, 1, 0, 1, 0, 1, 0, 0, 0, 0, 0, 0, 0, 1, 0, 1, 1, 1, 0, 1, 1, 1, 1, 0, 1, 1, 0, 0, 1, 1, 1, 1, 1, 1, 1, 1, 0, 0, 1, 0, 1, 1, 0, 1, 0, 1, 1, 0, 1, 1, 1, 0, 1, 0, 1, 0, 1, 0, 1, 1, 1, 0, 0, 1, 0, 0, 1, 1, 0, 1, 1, 0, 1, 0, 0, 1, 1, 0, 0, 1, 1, 0, 1, 0, 0, 0, 1, 1, 1, 0, 1, 1, 0, 1, 1, 0, 0, 0, 1, 0, 0, 0, 1, 0, 0, 1, 1, 1, 1, 0, 1, 0, 0, 1, 0, 0, 1, 0, 0, 0, 0, 1, 1, 1, 1, 0, 0, 0, 1, 0, 1, 0, 0, 1, 1, 1, 0, 0, 0, 1, 1, 1, 1). A discussion of these maximum length sequences and a table of possible algorithms for generating them can be found in Refs. 3-5.

Maximum length sequences have very good autocorrelation properties. If the elements of the sequence are multiplied by 2 and 1 is subtracted, a new sequence containing 1s and -1s will be obtained. The cross correlation between this new sequence and a shifted version of that sequence is $N$ when they are in sync and -1 if they are not, where $N$ is the length of the sequence. There are several ways of computing the correlation. One is by computing the quantity

$$R[n] = \sum_{n=0}^{N-1} A[m]B[m+n], \quad (3)$$

where $A$ and $B$ are two sequences of length $N = 2^k - 1$ and $k$ is the order of the sequence.

Computing the cross correlation directly using Eq. (3) works but requires the order of $N^2$ steps if all $N$ values of the sequence are computed. A faster way is to use Fourier transforms. In this case

$$R[n] = FFT^{-1}\left[FFT\left[A\right]^{*} FFT[B]\right], \quad (4)$$

where * indicates the complex conjugate and FFT is the Fast Fourier transform of the sequence. This way requires only the order of $N \log N$ steps versus $N^2$ steps to compute Eq. (3) for all $0 \le n \le N-1$. As an example we consider the sequence after Eq. (2) and compute the cross correlation between it and that same sequence shifted 128 places to the right. A plot of the cross correlation is shown in Fig. 1, which demonstrates the excellent delta function-like pulse profile of the resulting cross-correlation.

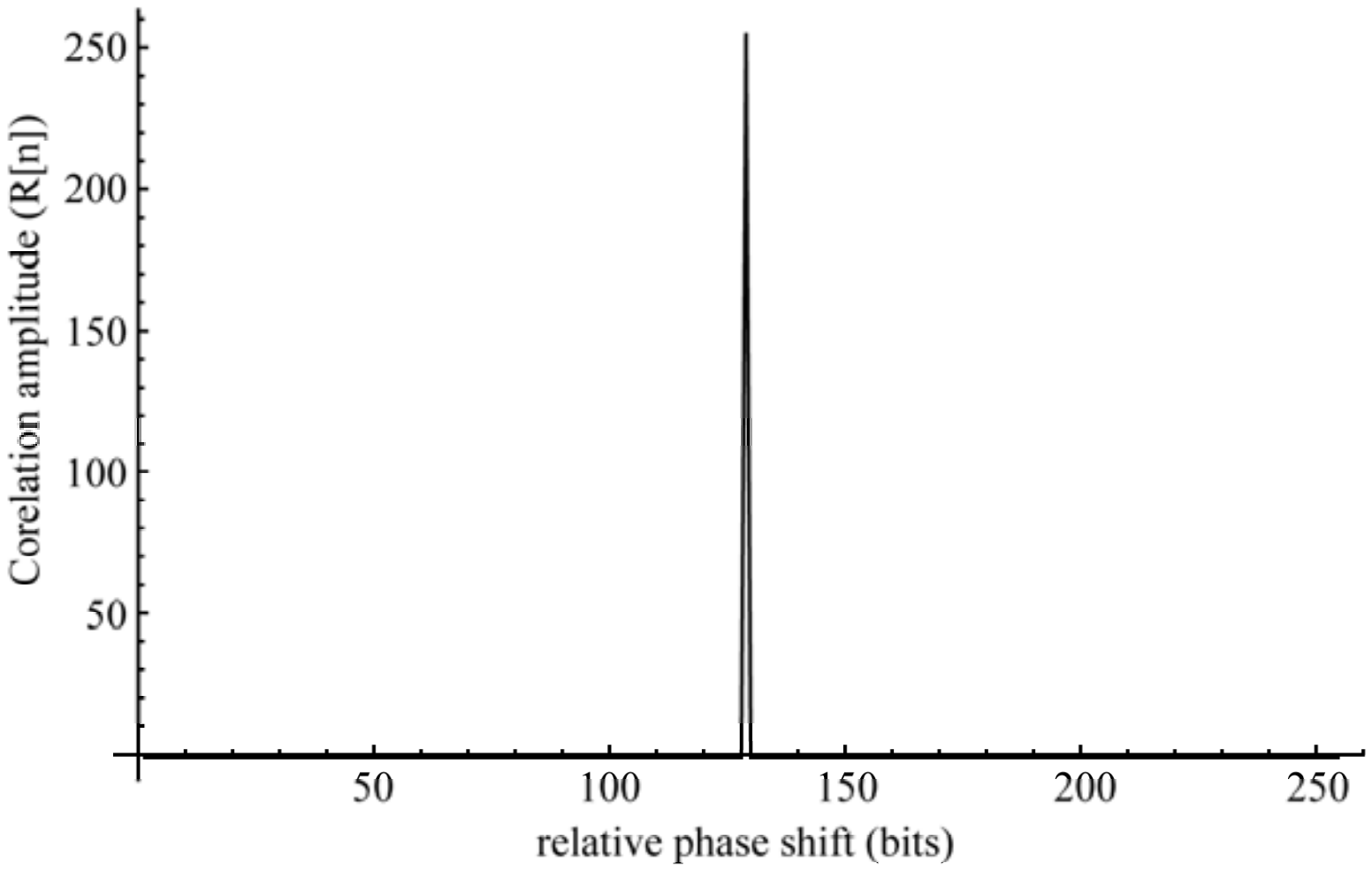

Fig. 1. Cross correlation of code and shifted code computed with a FFT.

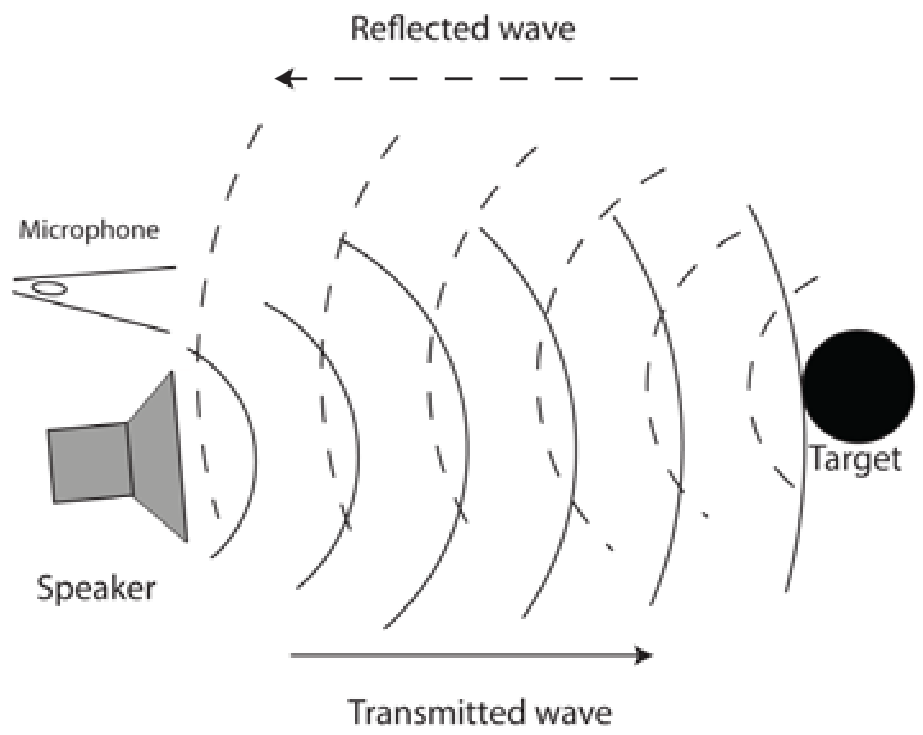

Fig 2. Sound waves reflecting off a hard target.

## III. RANGE AND RESOLUTION

In the ideal case we transmit the maximum length sequence without distortion through the air and let the transmitted sound wave bounce off a hard object as in Fig. 2. As mentioned, the distance of a pulse sonar is $d = c\ \tau / 2$, where $\tau$ is the time of flight to the target and back. If we use a pseudorandom sequence, then $\tau = n / bitrate$, where n is the number of sequence bits to the target and back and $\tau$ represents the time it takes for n bits of the sequence to reach the target and back. Therefore,

$$d = \frac{c}{2bitrate} n, \tag{5}$$

where *c* is the speed of sound and bitrate is the bit rate of the pseudorandom sequence. The resolution, which is defined by the pulse width, is the distance d divided by the number of bits because the pulse is 1 bit wide. The result is

$$r = \frac{c}{2bitrate}, \tag{6}$$

where *r* is the resolution. If an object is farther than the maximum number of bits of the sequence, then the cross correlation defined by Eqs. (3) and (4) wraps around. As a result there is a tradeoff between the bit rate and sequence length or resolution and range. High bit rates give good resolution but can restrict the maximum range for a particular sequence length. Conversely, there is a limit to what bit rate we choose based on the speed of the digitizer and other factors.

In any practical system using sound hardware we are restricted to specific sample rates such as 48k samples/s. To have flexibility in the sequence bit rate we must over sample the sequence by a specific amount. By over sampling we mean digitizing the sequence at a rate higher than one sample per sequence bit. If we over sample each sequence element by a factor of *M*, the distance between points after correlation is

$$r' = \frac{c}{2M\ bitrate} = \frac{c}{2samplerate}, \tag{7}$$

where $r'$ is the data resolution. The true resolution is the smallest distance we can resolve between any two peaks which is determined by the width of the peak and which is always given by Eq. (6) whether we over sample the sequence or not. However, the distance to the target in terms of sample points $p$ becomes

$$d = p\ r', \tag{8}$$

where $0 < p \leq M\ N$, and thus the maximum target distance does not change with over sampling.

If the over sampled sequences elements are multiplied by two and one is subtracted, the cross correlation properties are that the cross correlation is $MN$ when they are in phase and -$M$ when they are out of phase.

## IV. HARDWARE IMPLEMENTATION

For practical reasons, it is much better to put the pseudorandom sequence on a audio carrier signal. It is possible to transmit the code directly, but because the power band of the pseudorandom sequence is centered about direct current (DC), and the speakers are not capable of transmitting DC, the result will be a severely distorted sequence, which would degrade the autocorrelation properties. By using a carrier frequency that is centered in the bandwidth of the speaker/microphone combination, we can better take advantage of the bandwidth available. One way to measure the bandwidth is to play white noise through the speaker while simultaneously recording it with a microphone. By taking the Fourier transform we can see the frequency response of the system. A smooth curve can be generated by averaging many frames of the resulting power spectrum.

In our implementation we use amplitude modulation, which is easier because all we have to do is turn the carrier on and off according to the ones and zeroes of the code, like a random Morse code. If implemented on a computer, this procedure can be as simple as creating a particular sound file and playing it through a computer speaker. Demodulating the signal can also be done in software using a Hilbert transform. The signal processing is the difficult part. The easy part is putting the hardware together. In the simplest implementation all that is needed is a computer with a sound card and line level microphone input, a computer speaker, and a directional microphone. The reason we need a line level sound card is because we loop one half of the stereo output back to the microphone input (see Fig. 3) and use that as a reference for the cross correlation. Most personal computers other than a Macintosh have a powered microphone level input that uses different levels than standard line level computer equipment. Therefore, if you interface a standard computer microphone input to the stereo out for the loopback cable, the result will be a distorted signal because the microphone input will be saturated. If a line level stereo input is not available, an inexpensive USB sound card such as the Griffin iMic is an obvious solution. They are designed for a Macintosh, but work on any personal computer hardware with a USB port, without special drivers.

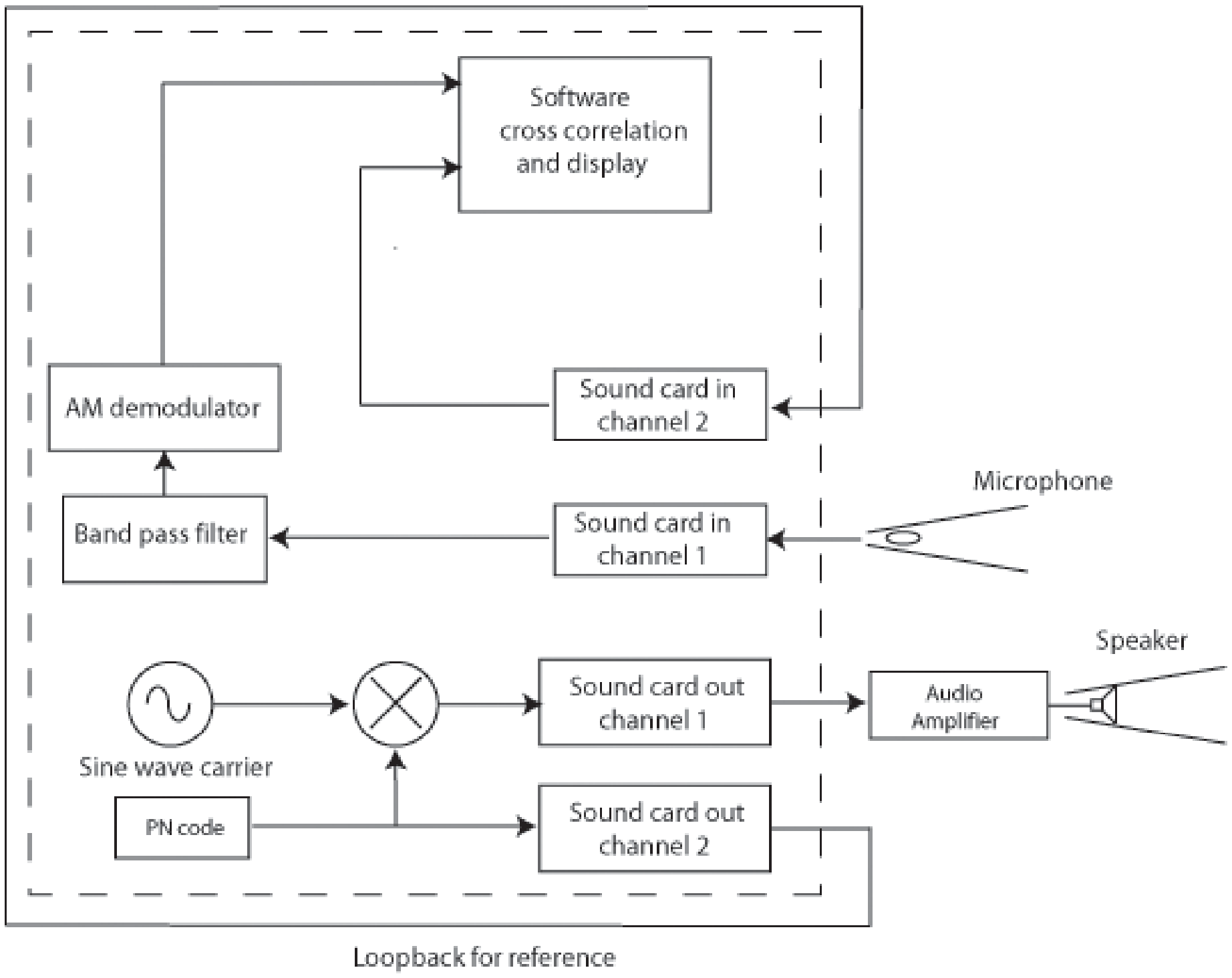

Figure 3. A simple PC system for measuring range. Enclosed dashed line shows which functions are either performed in software or internal to the PC.

In our initial experiments we used am inexpensive microphone embedded in an oil funnel to make it directional. The microphone was coupled with am inexpensive computer speaker. This setup makes a very impressive demonstration when coupled with a real time software display such as Labview. An improvement to this design can be done using a high quality studio amp, and high frequency speakers. We experimented with many of these, and found just about any tweeter designed to operate up to 30 kHz was sufficient. Models that operate up to 120 kHz are available. There are also cheap models that screw into inexpensive exponential horns.

Studio amplifiers have a much wider bandwidth than normal consumer stereo equipment. They are not particularly expensive, especially when purchased used.

Cross talk is generally not an issue. Putting the microphone inside a funnel or exponential horn is sufficient isolation to eliminate most (but not necessarily all) of the crosstalk. The amount left over is not an issue either because the residual crosstalk results in a peak near zero distance after correlating the outgoing and incoming signals, which is typically far from the reflected peaks. By analogy a pulse system can also have crosstalk, but it isn't an issue because the received pulses occur at different times.

We used a low end system and a more expensive system. The low end system consisted of an old Macbook Pro with Windows XP running under Boot Camp. A Windows laptop may be substituted in either system. We also used a cheap computer speaker/amplifier, the internal sound card of the Macbook Pro, a microphone (Radio Shack model AVL516), and an oil funnel. We purchased ours for about $2 and cut off the narrow end and inserted the microphone. The microphone preamplifier was a MXL iBooster unit for $40, which has the advantage of small size. Higher end sound cards sometimes have a preamp built in and 48 V for powered mics. Labview software using Christian Zeitnitz' WaveIO sound card interface was also used.[6]

Our more expensive system used a Samson Servo 201a studio amplifier (with 100 W/channel and a flat response up to 65 kHz), a M-Audio Firewire 1814 sound card with 192 kHz at 24 bits sample rate with a built in preamplifier and 1394a Firewire interface, and a Selenium HL14-25 exponential horn. We embedded the Behringer ECM8000 measurement microphone using styrofoam in the exponential horn. This microphone requires 48 V, which is provided by the sound card. It has a flat response up to 20 kHz. We also used is the Goldwood GT-400CD Bullet Piezo horn driver rated to 30 kHz. There is enough sensitivity to measure to much higher frequencies. We did so up to 45 kHz, albeit with a reduced sensitivity. The same software was used.

## V. REMOVING THE CARRIER

Because we used a carrier to transmit the sequence, we must remove the carrier before we do the cross correlation. An effective way to do this is with a Hilbert transform defined by

$$H\left[S(t)\right]=\frac{1}{\pi}P\int_{-\infty}^{\infty}\frac{1}{t-t'}S(t')\,dt', \tag{9}$$

where $H$ is the Hilbert transform, $S$ is the signal, and P is the Cauchy principal value. Hilbert transforms take a sinusoidal signal and shift its magnitude by $\pi/2$ radians such that

$$\left|H\left[\sin(\omega t+\phi)\right]\right|=\left|\cos(\omega t+\phi)\right|. \tag{10}$$

By summing the square of the original signal and the [square of the Hilbert transform of the signal we remove the carrier and are left with the pseudorandom code modulation. If the amplitude modulation is slow compared to the carrier,

$$\begin{aligned}&H\left[A(t)\sin(\omega t+\phi)\right]^2+\left[A(t)\sin(\omega t+\phi)\right]^2\\&\approx\left[A(t)\cos(\omega t+\phi)\right]^2+\left[A(t)\sin(\omega t+\phi)\right]^2=A(t)^2\end{aligned} \tag{11}$$

where $A(t)$ is the amplitude modulation, $\omega$ is the angular frequency, $t$ is the time, and $\phi$ is the phase. From a practical standpoint it is inefficient to compute Eq. (9) directly. Instead, it is better to express the Hilbert transform in terms of the Fourier transform. We define the Fourier transform and its inverse as

$$\begin{aligned}s(f)&=F\left[S(t)\right]=\int_{-\infty}^{\infty}S(t')\exp(-2\pi ift')dt',\\S(t)&=F^{-1}\left[s(f)\right]=\int_{-\infty}^{\infty}s(f')\exp(2\pi if't)df',\end{aligned} \tag{12}$$

where $f$ is the frequency and s is the Fourier transform of the signal S. From Eq. (12) we find

$$F\left[H\left[S(t)\right]\right] = -i \ \mathrm{sign}(f) \ s(f), \tag{13}$$

where the sign function has the usual meaning

$$\begin{aligned} \mathrm{sign}(x) &= 1, x > 0, \\ \mathrm{sign}(x) &= 0, x = 0, \\ \mathrm{sign}(x) &= -1, x < 0, \end{aligned} \tag{14}$$

so that

$$H\left[S(t)\right] = F^{-1}\left[-i \ \mathrm{sign}(f) \ s(f)\right]. \tag{15}$$

For our purposes the discrete Hilbert transform is more appropriate because we are sampling the signal discretely. We define the discrete Fourier transform as,

$$\begin{aligned} s(r) &= DFT\left[S(q)\right] = \frac{1}{\sqrt{N}} \sum_{q=0}^{N-1} S(q) \exp(-2\pi irq / N), \\ S(q) &= DFT^{-1}\left[s(r)\right] = \frac{1}{\sqrt{N}} \sum_{r=0}^{N-1} s(r) \exp(2\pi irq / N). \end{aligned} \tag{16}$$

The discrete Hilbert transform becomes

$$DHT\left[S(q)\right] = DFT^{-1}\left[\xi(r) \ s(r)\right], \tag{17}$$

where

$$\begin{aligned} &\xi(r) = -i \ sign(r - N/2), 1 \le r \le N-1, \\ &\xi(r) = 0, \ r = 0. \end{aligned} \tag{18}$$

Figure 4 shows a sample modulation/demodulation of a pseudorandom signal.

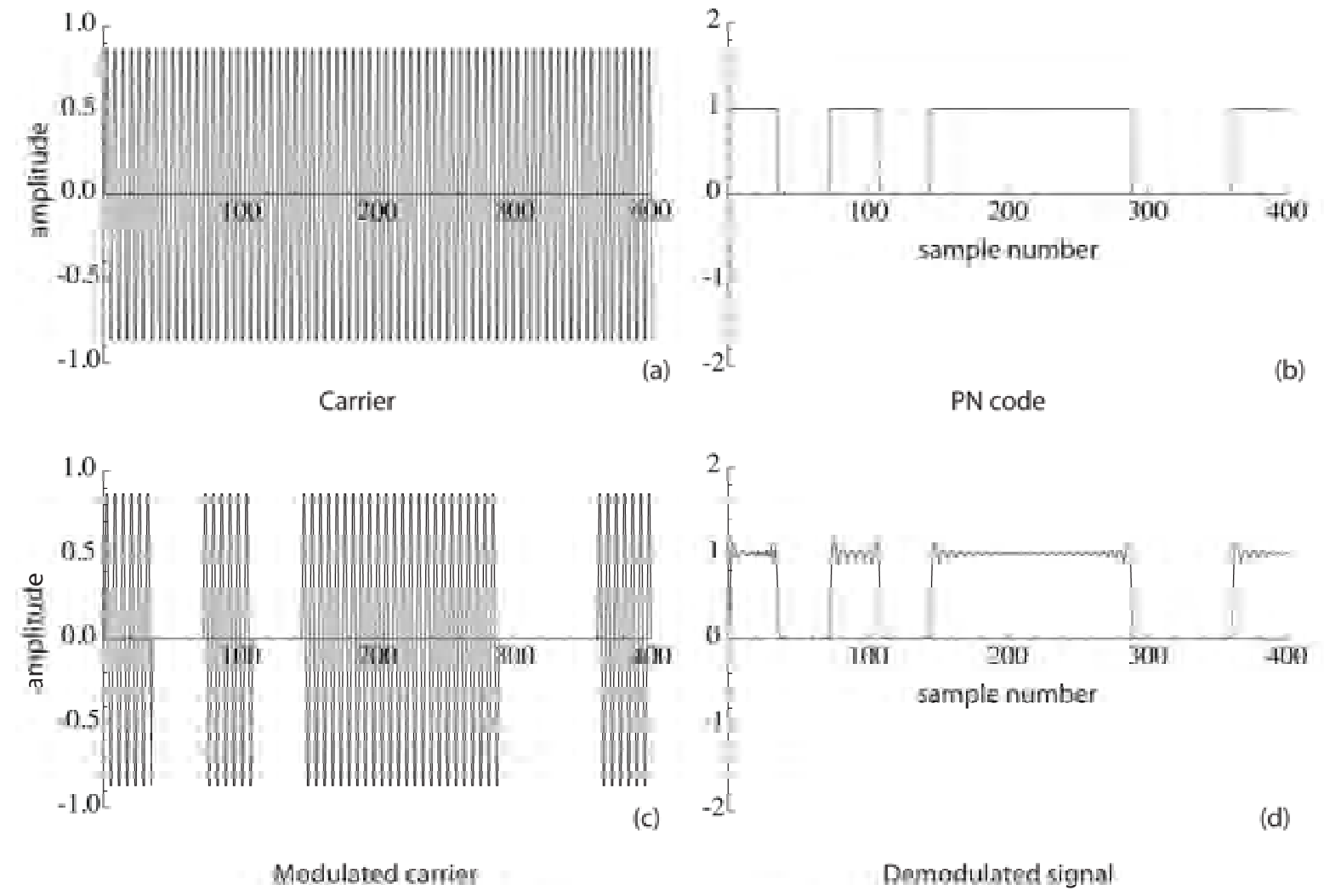

Figure 4. Carrier demodulation using the Hilbert transform.

## VI. DETERMINATION OF DISTANCE

Determining the distance to the target is a simple task using the implementation we have described. Once the signal is demodulated we have to compute the cross correlation between the left and right stereo channels over the length of a single frame, where each frame represents the repeat modulation pattern from beginning to end. Each frame has a length *MN*, and it does not matter if the correlation starts in the middle of a frame because the correlation properties for the maximum length sequences are the same no matter where in the frame we start as long as each repeating frame is identical. The beauty of this approach is that by digitizing the outgoing sequence and incoming data simultaneously, we can determine the approximate distance to the target by comparing the relative phase of the left and right stereo channels using the cross correlation. There is no complicated book keeping involved. Once this correlation is done, all we have to do is make a table of amplitude versus sample number multiplied by r' given by Eq. (7) as in Eq. (8). The resolution error of the measurement is determined by the true resolution given by Eq. (6) such that

$$\delta d = \frac{c}{2 samplerate}\delta p = \frac{c}{2 samplerate} M = \frac{c}{2 bitrate}, \tag{19}$$

because we are measuring the distance with a pulse of a specific width such that the pulse width determines the resolution error. The fractional error due to resolution can be determined by dividing the uncertainty in the distance measurement by the magnitude of the distance. The uncertainty in the distance is the width of the pulse so that

$$\frac{\delta d}{d} = \frac{1}{d}\frac{c}{2 samplerate}\delta p = \frac{\delta p}{p} = \frac{M}{p} = \frac{1}{n}, \tag{20}$$

where *n* is the number of code bits to the target and $\delta p$ is the uncertainty in the number of samples to the target. Equation (20) implies that the farther the target, the smaller the fractional error. We can make the code as long as we wish.

Other errors include knowing where the measurement starts and where the sound wave hits which depend on the experimental setup. If the measurement distance is long enough, these errors are less important in terms of the percentage of error. For shorter distances we could do a differential measurement to remove the errors, but doing a differential measurement makes sense only if the resolution error is small compared to the bias error, which can be achieved by increasing the bit rate and also the code length if longer distances are also required.

## VII. SOFTWARE IMPLEMENTATION

There are a number of possibilities for implementing the block diagram represented by Fig. 3 in software. A simple method is to create custom sound files and play them at the appropriate bit rate in a program such as Apple's Quicktime player. If the sound is simultaneously recorded in a program such as Audacity[7] with the wiring shown in Fig. 3, we can compute the range profile by post processing the data (storing the data on the hard drive for later processing rather than real time processing) by demodulating the signal, then performing the required filtering and cross correlations between stereo channels. Although awkward, this method is sufficiently simple that students could use it to obtain results with a minimum of programming – especially if the sound files were created for them. Mathematica has the intrinsic ability to create and read sound files so that it is a possible platform for processing.[8] Other possibilities include Matlab/Octave, both of which have freeware sound card interfaces.[9]

We have created a real time system in Labview. It is a simple language for students to learn who don't know how to program. To take full advantage of the capabilities of the computer sound hardware, we recommend using the WaveIO sound card interface developed by Christian Zeitnitz.[6] By modifying the settings, we can potentially sample up to 192 kHz using this software, which makes an ultrasonic system a real possibility. A unique feature of this software is that we can set the sample rate to a non-standard rate such as 100 kHz for instance. A sample output of this software is shown in Fig. 5.

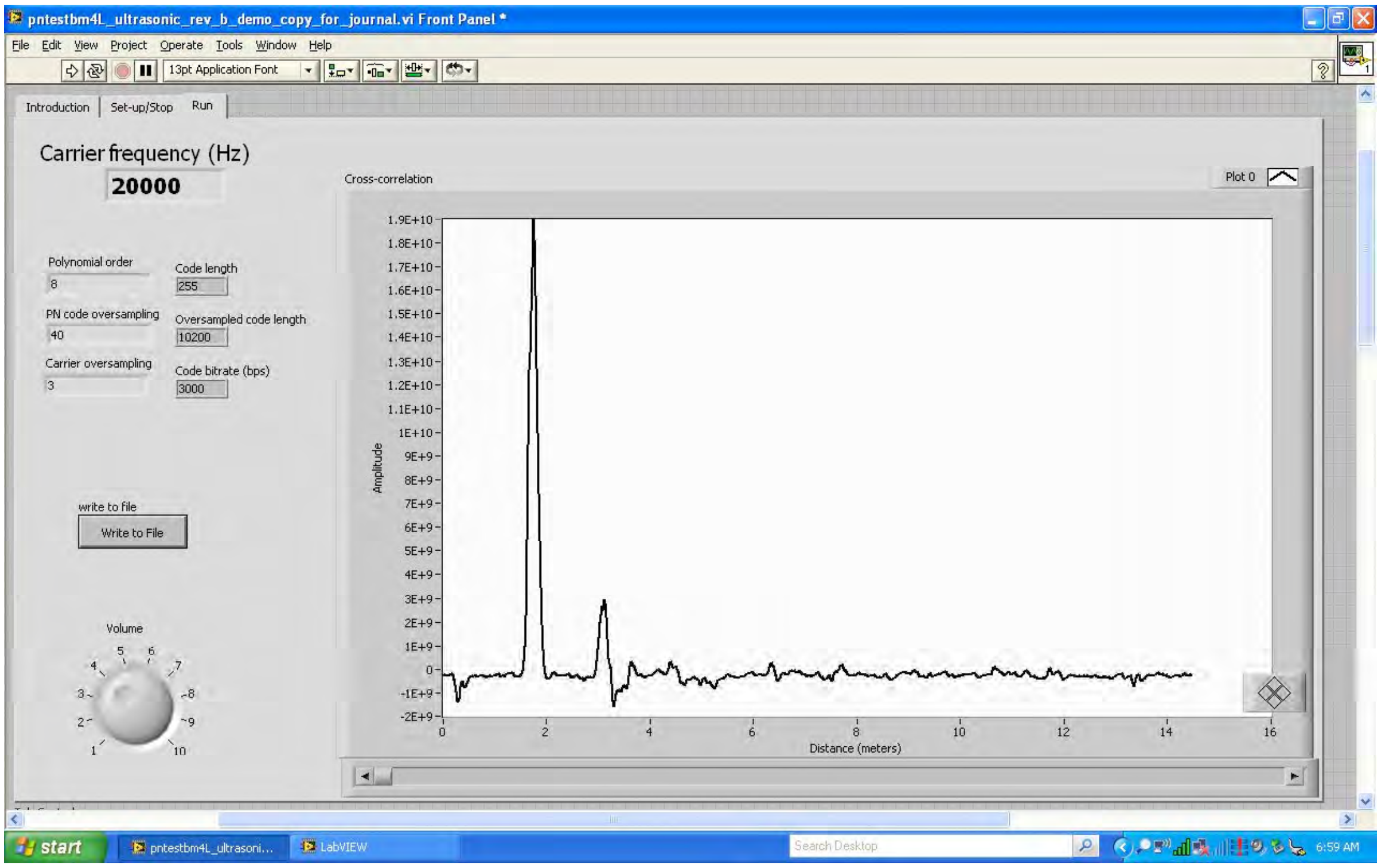

Figure 5. Sample Labview program output for sonic/ultrasonic ranging system. The target was a large metal file cabinet. The large return was from the front of the cabinet and the second smaller return was from a set of glass doors behind the cabinet.

We have also developed a dual channel system capable of demonstrating differential absorption measurements using pseudorandom codes. The details can be found in Ref. 10.

## VIII. SYSTEM DEMONSTRATION

We first used the more expensive system (see Fig. 6) to make some rough measurements using sample rate: 120 kHz @ 16 bits; code length:255; bit rate: 3 kHz; code oversampling: 40; carrier frequency: 20 kHz; and bandpass filter range: 17-23 kHz. We measured the distance to the target, which was a large file cabinet placed in the center of a long hallway, as 13.90 m using a portable ultrasonic range finder, which roughly corresponded to the distance reported by the system.

The real value of this system is as a demonstrate some of the concepts of remote sensing. Because the amplifier is capable of an output power of up to 100 W and the sensitivity of the electronics and sound card is high, this system could potentially be used to detect objects at distances similar to a portable radar or sonar system. The system could be improved by using a large parabolic reflector dish to aid in detecting weak signals to the microphone. Because the sound card can be set to use 24 bit data resolution and has a volume control, there is a very large dynamic range to perform such measurements. For long distance applications we would typically use a longer code and could also reduce the bit rate. In the Labview software the code length is one of the parameters.

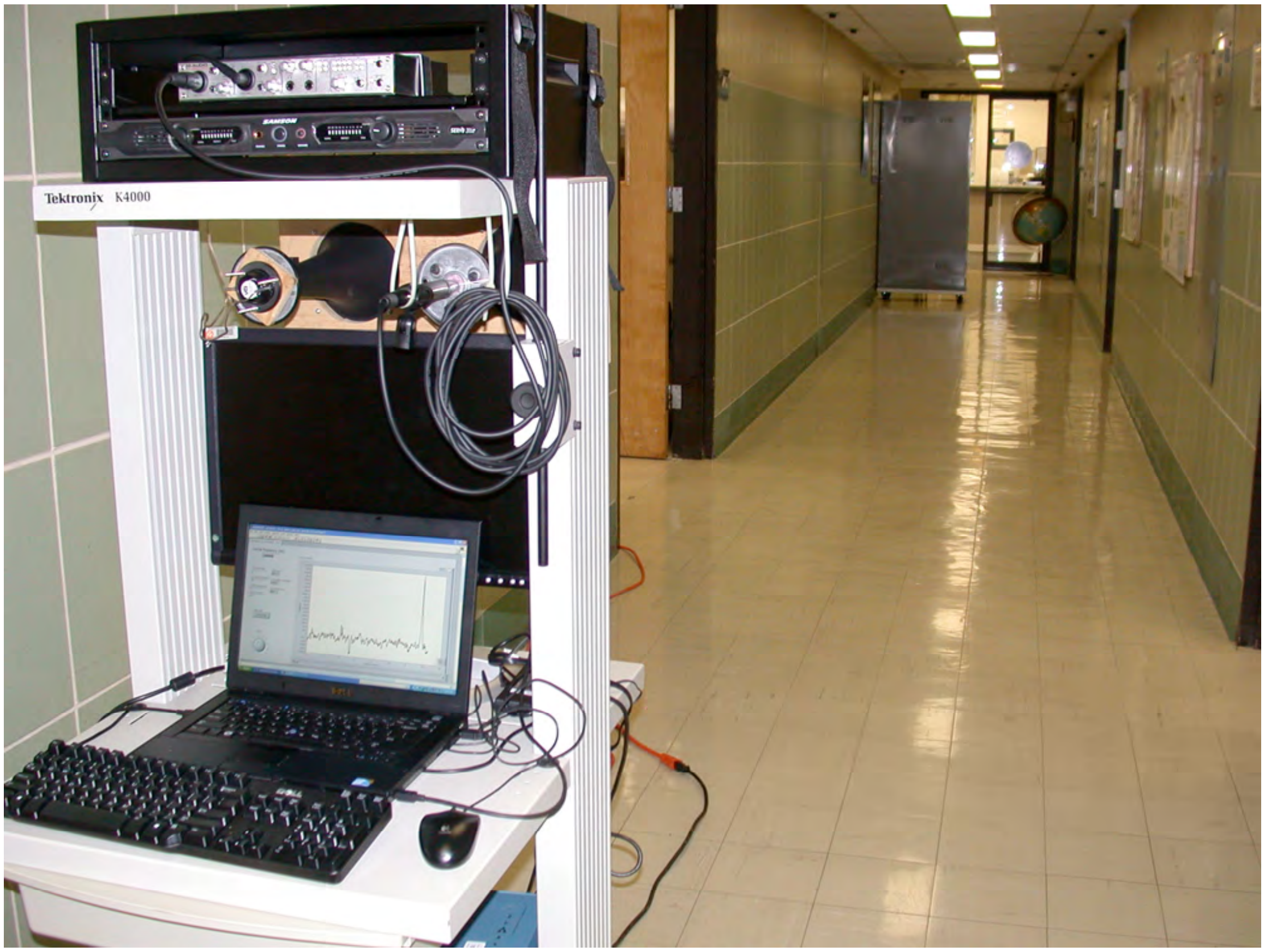

Figure 6. Experimental setup for ultrasonic ranging system capable of transmitting a carrier up to 45 kHz.

For the less expensive end experiment the software was set to the following settings: sample rate: 44.1 kHz @ 16 bits; code length: 255; bit rate: 4.9 kHz; code oversampling: 9; carrier frequency: 11.025 kHz; and bandpass filter range: 6.125-15.925 kHz. Figure 7 shows a screen shot of a measurement where the speaker and microphone/funnel were pointed at the ceiling. The screen was frozen so that we could move the device to a location with more light so that we could take a picture of it. The distance measured was about 1.5 m, which roughly corresponds to the actual distance. This system makes a perfect experiment for students interested in simulating a radar or lidar system in the home. In the Labview implementation we can see in real time the peak move back and forth as we walk closer or farther from a reflector such as a wall or ceiling.

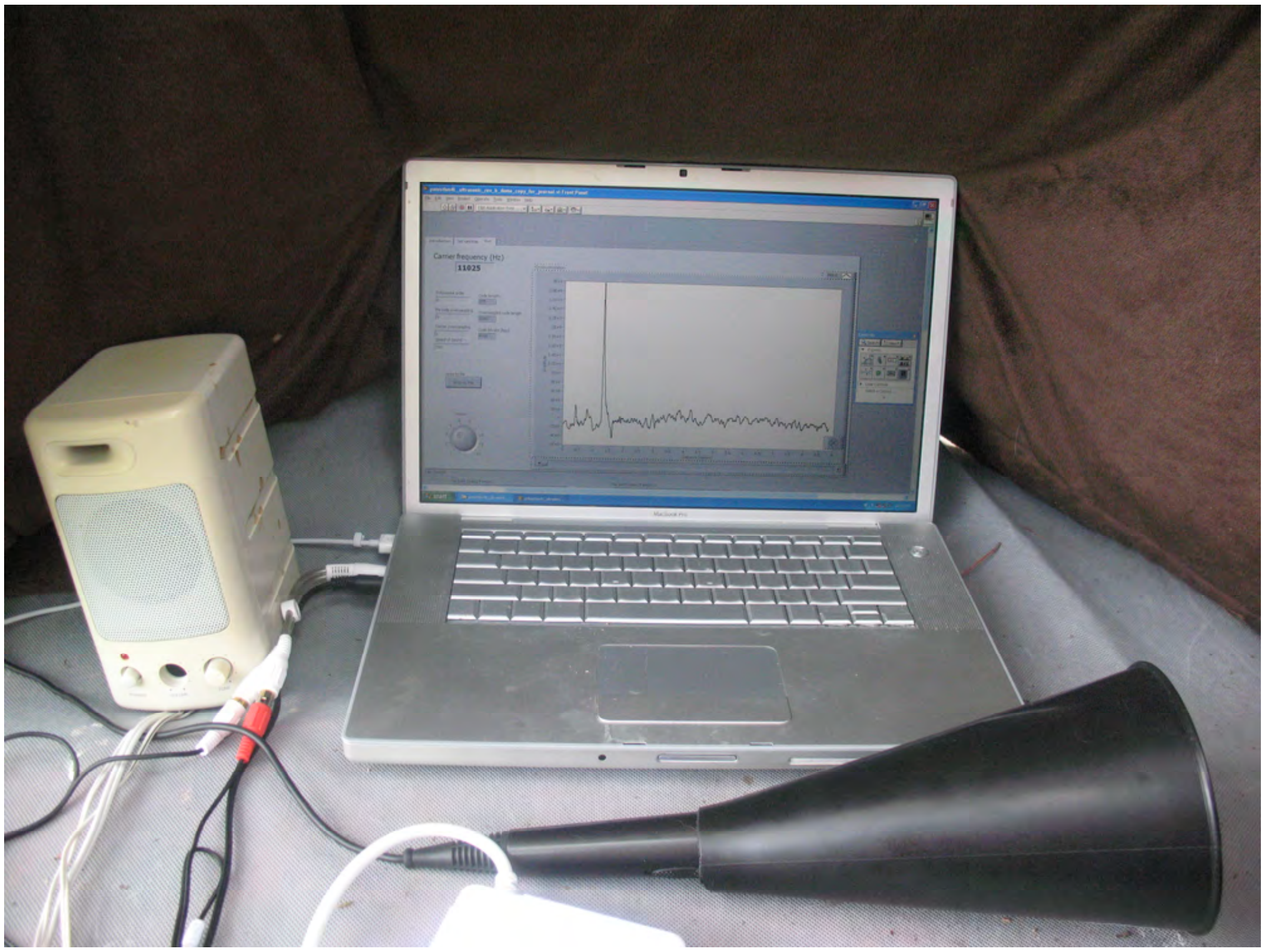

Figure 7. Home experiment configuration using a PC speaker/amplifier and mic embedded in an oil funnel.

## IX. DISCUSSION

Our results show how to build a simple sounding system with virtually no specialized hardware outside of what most people have available at home and in the lab. We also show how to improve the system using higher quality equipment. Although other modulation schemes are possible, this one produces superior results due to the very good autocorrelation properties of ML sequences. From a hardware standpoint, this is something almost anyone could put together, reducing the problem to one of software development.

## REFERENCES

1. N. Takeuchi, N. Sugimoto, H. Baba, and K. Sakurai, "Random modulation CW lidar," Appl. Optics **22**(9), 1382-1386 (1983).

2. Nobuo Takeuchi, Hiroshi Baba, Katsumi Sakurai, and Toshiyuki Ueno, "Diode-laser random-modulation CW lidar," Appl. Optics **25**(1), 63-67 (1986).

3. F. J. MacWilliams and N. J. A. Sloane, "Pseudo-random sequences and arrays," Proc. IEEE **64**(12) 1715-1729 (1976).

4. D. V. Sarwate and M. B. Pursley, "Crosscorrelation properties of pseudorandom and related sequences," Proc. IEEE **68**(5), 593- 619 (1980).

5. "Linear feedback shift registers implementation, M-sequence properties, feedback tables new wave instruments"

6. Christian Zeitnitz, "WaveIO: A soundcard interface for Labview"

7. Audacity.

8. Joel F. Campbell, N. Prasad, Michael Flood, and Fenton Harrison, “Experiments in Mathematica with spread spectrum SONAR,” Wolfram Technology Conference, October 14, 2010.

9. Playrec is an example of a sound card interface that works with both Octave and Matlab.

10. Joel Campbell, Narasimha S. Prasad, Michael Flood, and Wallace Harrison, “Pseudorandom noise code-based technique for cloud and aerosol discrimination applications”, Proc. SPIE **8037**, 80370L (2011).